\documentclass{aa}  
\usepackage{hyperref}
\usepackage{graphicx,graphics,rotating,amssymb,amsmath}
\usepackage{amsmath}
\usepackage{physics}
\usepackage{amssymb}
\usepackage{xcolor}
\usepackage[utf8]{inputenc}
\usepackage{graphicx}
\usepackage{lipsum} 
\usepackage{pdfpages}

\usepackage{txfonts}
\newcommand{\Cloudy}{\textsc{Cloudy}}

\newcommand{\Starburst}{\textsc{Starburst 99}}

\newcommand{\Gasoline}{\textsc{Gasoline}}
\newcommand{\Ponos}{\textsc{Ponos}}

\def\kramsesrt{\textsc{Kramses-rt}}
\def\ramsesrt{\textsc{Ramses-rt}}
\def\romulus25{\textsc{Romulus25 }}

\def\krome{\textsc{Krome}}

\usepackage{lscape}
\usepackage{longtable} 
\usepackage{array}

\definecolor{orange}{HTML}{fc8e2d}

\newcommand{\rev}[1]{\textcolor{black}{#1}}

\begin{document} 

        \title{Constraining the physical properties of gas in high-z galaxies with far-infrared and submillimetre line ratios}
        
       \titlerunning{Constraining gas properties in high-z galaxies with FIR and submm line ratios}
  
\author{A. Schimek
          \inst{1}
          \and C. Cicone
          \inst{1}
          \and S. Shen
          \inst{1}
          \and D. Decataldo
          \inst{1}
          \and P. Klaassen
          \inst{2}
          \and L. Mayer
          \inst{3}
        }

   \institute{Institute of Theoretical Astrophysics, University of Oslo, PO Box 1029, Blindern 0315, Oslo, Norway \\ \email{alice.schimek@astro.uio.no}
   \and UK Astronomy Technology Centre, Royal Observatory Edinburgh, Blackford Hill, Edinburgh United Kingdom
   \and Department of Astrophysics, University of Zurich, Winterthurerstrasse 190, CH-8057 Zurich, Switzerland
    }

   \date{Received ; accepted }

 
  \abstract %
  {Optical emission line diagnostics, which are a common tool for constraining the properties of the interstellar medium (ISM) of galaxies, become progressively inaccessible at higher redshifts for ground-based facilities. Far-infrared (FIR) emission lines, which are redshifted into atmospheric windows that are accessible for ground-based submillimetre facilities, could provide ISM diagnostics alternative to optical emission lines. We investigated FIR line ratios involving [CII]~$\lambda 158 \mu$m, [OIII]~$\lambda 88 \mu$m, [OIII]~$\lambda 52 \mu$m, [NII]~$\lambda 122 \mu$m, and [NIII]~$\lambda 57 \mu$m using synthetic emission lines applied to a high-resolution (m$_{\rm gas}$= 883.4 M$_{\odot}$) cosmological zoom-in simulation, including radiative transfer post-processing with the code \kramsesrt~at z = 6.5. We find that the [CII]/[NII]122 ratio is sensitive to the temperature and density of photodissociation regions. It might therefore be a useful tool for tracing the properties of this gas phase in galaxies. We also find that [NII]/[NIII] is a good tracer of the temperature and that [OIII]52/[OIII]88 is a good tracer of the gas density of HII regions. Emission line ratios containing the [OIII]~$\lambda 88 \mu$m line are sensitive to high-velocity outflowing gas.}


\maketitle
%
\section{Introduction}
\label{Sec:Intro}

Far-infrared (FIR) emission lines are useful tools for the investigation of the interstellar medium (ISM) of galaxies because they are mostly unaffected by dust extinction \citep{Brauher08}. In recent years, FIR emission lines have been used to explore the gas properties of galaxies, such as the ionisation parameter, gas density, gas metallicity, gas temperature, and the radiation field \citep{Nagao11, Nagao13, Cormier15,Pereira17, Rigo18, Killi23, RP23, Kumari24}.  
At $z>2$ , several FIR lines are shifted into submillimetre (submm) atmospheric windows, where they can be observed by sensitive ground-based submm telescopes. This makes them attractive tools for characterising the ISM of high-redshift galaxies in the early stages of their evolution.

In \cite{Schimek23} we presented a line-modelling effort that included the [CII]~$\lambda 157 \mu$m, [CI](1-0)~$\lambda 609 \mu$m, [CI](2-1)~$\lambda370 \mu$m, CO(3-2)~$\lambda 866 \mu$m, and [OIII]~$\lambda 88 \mu$m lines. 
In this follow-up work, we adopt the same simulation and fiducial radiative transfer model as \cite{Schimek23} and expand it to include the additional transitions of [OIII]~$\lambda 52 \mu$m, [NII]~$\lambda 122 \mu$m, and [NIII]~$\lambda 57 \mu$m, which trace the warm gas found in HII regions \citep{Nagao11}. While in \cite{Schimek23} the main subject was to study the extended submm emission of the circumgalactic medium (CGM), we now focus on the inner part of the simulation.
Inspired by analytical modelling of these line ratios as seen in \cite{Nagao11} and \cite{Pereira17}, we focus on the change in the FIR emission line ratios as functions of ISM properties when applied to a full galaxy simulation. Examples of previously explored FIR emission line ratios are the [OIII]~$\lambda 88 \mu$m / [CII]~$\lambda 158 \mu$m ratio, which has been investigated in both local and high-redshift galaxies \citep{Vallini15, Inoue16, Carniani18,  Hashimoto19, Pallottini17, Pallottini19, Katz19,Harikane20, Arata20, Carniani20, Lupi2020,Vallini21,Fujimoto22, Katz22, Pallottini22, Witstok22, Kumari23}, and [OIII]~$\lambda 52 \mu$m / [OIII]~$\lambda 88 \mu$m, which has been shown to trace the gas density of ionised gas \citep{Dinerstein85,Rubin94, Brauher08, Nagao11, Nagao13, Pereira17, Rigo18, Killi23}. 
Our results are aimed to inform observations performed with current and future submm facilities, such as the Atacama Large Millimeter/submillimeter Array (ALMA), the Atacama Pathfinder Experiment (APEX), the Northern Extended Millimeter Array (NOEMA), and, in the future, the Atacama Large Aperture Submillimeter Telescope (AtLAST)\footnote{https://www.atlast.uio.no} \citep{Mroczkowski2024}. 

\section{Simulation and modelling}
\label{Sec:Sim}

\begin{figure}
   \includegraphics[clip=true,trim=2cm 0cm 0.0cm 0cm,scale=.25]{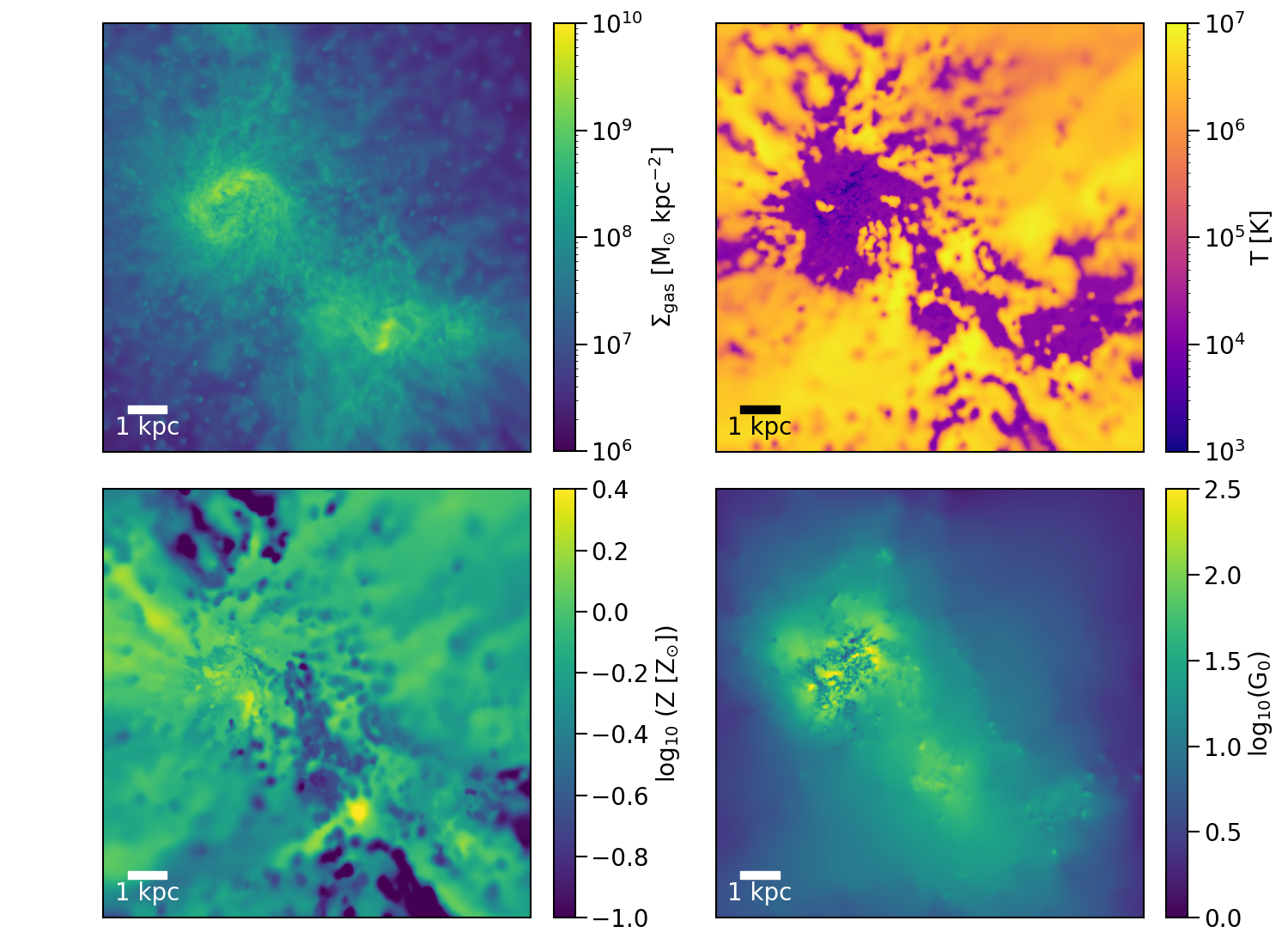} 
      \caption{Zoomed-in view of \Ponos, focusing on the main disc and the main merger. In the top row, the gas surface density is shown on the left and the gas temperature on the right. In the bottom row, the gas-phase metallicity is shown on the left and the FUV field on the right (for full-sized images of the simulation, see Fig.~1 in \cite{Schimek23}}).
         \label{fig:PonosMaps}
\end{figure}

In this study, we analyse the \Ponos~simulation \citep{Fiacconi17}, a high-resolution cosmological zoom-in simulation that was run with the smoothed particle hydrodynamics (SPH) code \Gasoline~\citep{Wadsley04}. The simulation has a gas mass resolution of m$_{\rm gas} = 883.4 $~M$_{\odot}$ and a stellar particle mass of m$_{*} = 0.4 ~\cdot$~m$_{\rm gas} = 353.4 $~M$_{\odot}$, and it uses the WMAP~7/9 cosmology ($\Omega_{m,0} = 0.272$, $\Omega_{\Lambda,0} = 0.728$, $\Omega_{b,0} = 0.0455$, $\sigma_{8} = 0.807$, n$_{s} = 0.961$, and H$_{0} = 70.2$~km~s$^{-1}$~Mpc$^{-1}$) \citep{Komatsu11, Hinshaw13}. 
\Ponos~is the progenitor of a massive galaxy, undergoing a merger (stellar mass merger ratio 1~:~2.7) in the analysed snapshot at $z=6.5$. The central galaxy has a stellar mass of M$_{*}= 2 \cdot 10^{9} $~M$_{\odot}$, a virial radius of R$_{\rm vir} = 21.18$~kpc, and a star formation rate, SFR, of $ 20$~M$_{*}$~yr$^{-1}$. These properties make \Ponos~a typical star-forming galaxy at $z=6.5$. Fig.~\ref{fig:PonosMaps} shows the gas density, temperature, metallicity, and far-ultraviolet (FUV) radiation field of the main disc and the main merger of \Ponos. Figures showing the entire halo were presented in \cite{Schimek23}.

We refer to \cite{Schimek23} for the details of emission-line modelling, but briefly outline our fiducial model here. We first converted the \Ponos~simulation into an adaptive mesh refinement (AMR) grid and then post-processed it with \kramsesrt~\citep{Pallottini19, Decataldo19, Decataldo20}, a customised version of \ramsesrt~\citep{Ramses, RamsesRT}, where a non-equilibrium chemical network generated via the package \krome~\citep{KROME} has been implemented. 
We used the \kramsesrt~ accurate scheme for radiative transport, accounting for photoreactions and gas self-shielding to compute the radiation in ten bins for each grid cell. We computed stellar spectra for each stellar particle with \Starburst~(SB99) \citep{SB99}, according to the age and the metallicity of the stellar particle, and we then rescaled it to its mass. In addition to stellar radiation, we included a uniform ultraviolet background (UVB) (\citealt{HM12} tables at $z = 6.49$) and attenuated it in each cell according to the column density of each chemical species. The simulation was further post-processed with \Cloudy~\citep{Cloudy}, creating a multi-dimensional grid for the line emission, based on the gas temperature, gas density, gas metallicity, and the incident FUV radiation field. \rev{We included the radiation field of the gas with  $G_{0}$, which corresponds to the ionising FUV radiation field (6 - 13.6 eV) in units of Habing ($1.6 \cdot 10^{-3}$ erg cm$^{-2}$ s$^{-1}$) \citep{Habing68}.}


To compute line luminosity ratios and line ratio maps from the post-processed simulation data, we applied a luminosity thresholds of L$_{\rm line} > 0.5$~L$_{\odot}$ per fluid element and $\Sigma_{\rm line} > 5$~L$_{\odot}$~kpc$^{-2}$ for the surface luminosity maps to the two lines involved in the ratio. These thresholds are very low and therefore do not affect the integrated ratio values of the halo.
The integrated line ratios for the entire halo and the different components can be seen in Table \ref{table:2}. We followed the same criteria to define the individual galaxy components as in \cite{Schimek23}, but in this study, we focus on the disc and merger components within the central 25~kpc because most lines in the CGM are faint. The applied threshold means that most of the far CGM is not included because the CGM components that are bright enough to satisfy the criteria are located close to the main disc and merger components within the central 25 kpc. 
We classified the gas into 'inflows' and 'outflows', in the same way as in \cite{Schimek23}. We selected `high-velocity' gas with an absolute $v>200$~km~s$^{-1}$ (higher than the rotational velocity of the main disc), and defined it as inflowing or outflowing by dividing it into gas that flowed towards and away from the main disc, respectively.

\section{Results}
\label{Sec:Results}

\begin{figure*}
\centering
   \includegraphics[width = 0.95\textwidth]{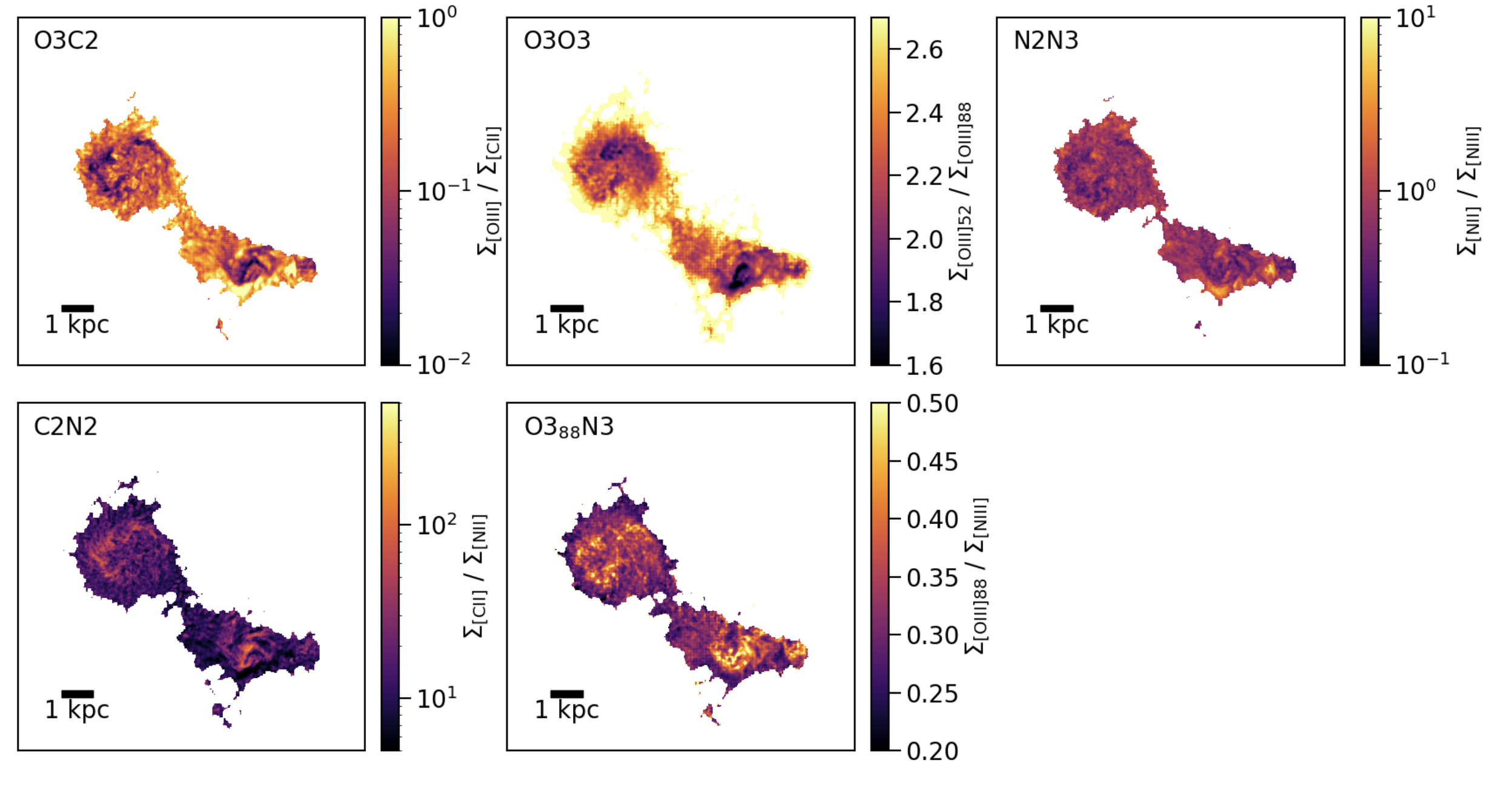} 
      \caption{Line ratio maps resulting from our modelling. Because of the surface brightness cuts applied to the two lines involved in the ratios ($\Sigma_{\rm line} > 5$~L$_{\odot}$~kpc$^{-2}$), only the central disc, major merger, and inner CGM components are probed. Only the central 25~kpc of the \Ponos~simulation halo is shown. }
         \label{fig:RatioMaps}
\end{figure*}

\begin{figure*}
\centering
   \includegraphics[width = 0.95\textwidth]{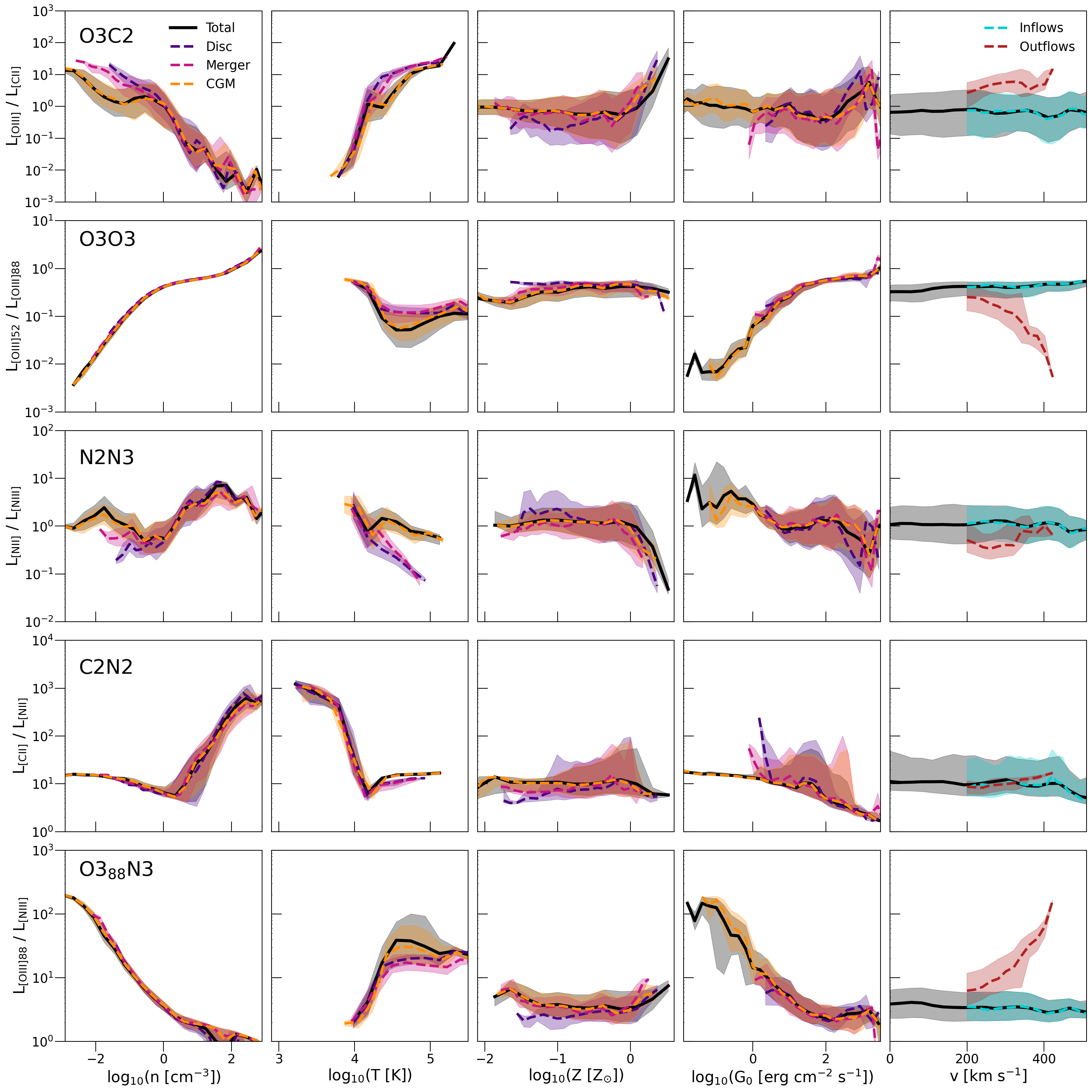} 
      \caption{Median of the line ratios plotted as a function of gas density, temperature, metallicity, UV field, and absolute velocity of the gas cells. The ratios are divided into the total halo (black), the disc (indigo), the merger components (magenta), and the CGM (orange), except for the absolute velocity panel, in which we divide the gas into inflowing (blue) and outflowing (red) gas. The shaded areas correspond to the 25th and 75th percentiles.}
         \label{fig:HistRatio}
\end{figure*}

\begin{table}
\caption{Integrated line ratio values of \Ponos for the different galaxy components.}             
\label{table:2}      
\centering                          
\begin{tabular}{l r r r r}        
\hline
\hline
  Line ratio          & Total     & Disc  & Merger    & CGM\\ \hline 
            
L$_{\rm [OIII]88}$/L$_{\rm [CII]158}$        & 0.17      & 0.13  & 0.18      & 0.39\\

L$_{\rm [OIII]52}$/L$_{\rm [OIII]88}$        & 0.36      & 0.43  & 0.45      & 0.13\\

L$_{\rm [NII]122}$/L$_{\rm [NIII]57}$        & 1.37      & 1.36  & 1.29      & 1.73\\

L$_{\rm [CII]158}$/L$_{\rm [NII]122}$        & 16.92     & 18.62 & 14.56     & 15.15\\


L$_{\rm [OIII]88}$/L$_{\rm [NIII]57}$        & 4.01      & 3.40  & 3.39      & 10.28\\ \hline


\end{tabular}
\end{table}

Our results are shown in Fig.~\ref{fig:RatioMaps}, displaying emission line ratio maps, and Fig.~\ref{fig:HistRatio}, showing the ratios as a function of gas density, temperature, metallicity, incident FUV radiation field, and absolute velocities of the gas cells. In the following, we comment on each line ratio separately. 


\subsection{O3C2}

The [OIII]~$\lambda 88 \mu$m / [CII]~$\lambda 158 \mu$m (O3C2) ratio is of great interest because the [CII] and the [OIII] emission lines are bright coolants of the ISM in the FIR and can be observed with ground-based telescopes from a high-altitude site at $z\gtrsim1$ and $z\gtrsim3$, respectively. 
[CII] is one of the brightest FIR emission lines in star-forming (SF) galaxies. It accounts for up to 0.1 - 1\% of the total FIR emission \citep{Stacey91} and traces both ionised and neutral gas because of the low excitation energy of carbon (11.26 eV, which is lower than hydrogen with 13.6 eV). [OIII] has a high ionisation energy of 35 eV and traces hot and diffuse gas in HII regions and hot outflows. 
The O3C2 ratio has been investigated in both observational and theoretical studies of high- \rev{and low-redshift} galaxies \citep{Vallini15, Inoue16, Carniani18,  Hashimoto19, Pallottini17, Pallottini19, Katz19,Harikane20, Arata20, Carniani20, Lupi2020,Vallini21,Fujimoto22, Katz22, Pallottini22, Witstok22, Kumari23, RP23}. 
As the [CII] and [OIII]88$\mu$m lines trace different gas phases within the ISM, the analysis of the emission line ratio can give information about the physical conditions of the ISM, such as the PDR covering fraction, gas densities, and temperatures.

In \cite{Schimek23} we showed that [CII] emission is more extended than [OIII], and while the respective line luminosities fall within the scatter of observational data, \Ponos~(total \rev{O3C2} ratio of 0.17) is more aligned with ratios in the local Universe than with high-redshift galaxies, which have observed ratios between 1 - 10 \citep{Harikane20}. \rev{We note that local dwarf galaxies, which are often used as analogues of early galaxies, appear to share some of these extreme conditions, with [CII] being more extended than [OIII] \citep{Cormier15} and O3C2 ratios exceeding unity \citep{Kumari24}}. 
One of the reasons for the enhanced ratios measured at high-redshift could lie in abundance ratio differences that are due to an early stage of chemical evolution (see Nyhagen et al. in prep).
The O3C2 map reported in the upper left panel of Fig.~\ref{fig:RatioMaps} shows that the ratio is \rev{significantly} enhanced in \rev{spatially compact areas in the disc, which correspond to hot low-density and high-metallicity regions (see Fig.\ref{fig:PonosMaps})}. 

\rev{In Fig.~\ref{fig:HistRatio}, where the first row corresponds to the O3C2 ratio, we} find that the O3C2 ratio is most sensitive to the gas density and temperature \rev{because it varies} by four orders of magnitude. It  \rev{is low in} high-density (n~$\sim 10^{3}$~cm$^{-3}$) and low-temperature gas (T~$\sim 8 \cdot 10^{3}$~K) \rev{and high in} low-density (n~$\sim 5 \cdot 10^{-3}$~cm$^{-3}$) and high-temperature gas (T~$\sim 5 \cdot 10^{5}$~K). 
\rev{We find O3C2 ratio values of $\leq$1 for typical values of HII regions (T$\sim$ 7$\cdot$10$^{3}$ - 1.5$\cdot$10$^{4}$~K, n$\sim$ 10 - 10$^{2}$~cm$^{-3}$, \cite{Osterbrock06}). The highest ratios ($\geq $10) are found in gas with typical densities and temperatures of SNe remnants (T$\sim$ 10$^{5.7}$~K, n$\sim$ 10$^{-2.5}$~cm$^{-3}$; \cite{McKee77}).}
The ratio shows little scatter as a function of the density and temperature, and the ISM components of the disc and the merger and the CGM are similar, probably because this analysis, as explained in Section\ref{Sec:Sim}, is not sensitive to the most extended CGM gas. 
The dependence of O3C2 on the radiation field seems not very significant up to log$_{10}$(G$_0$)~$\sim$~2, above which the ratio increases \rev{as the strong radiation can ionise [OIII], while carbon is doubly ionised, which decreases the [CII] emission}.
\rev{As we assumed uniform solar abundance ratios for the gas in our \Cloudy~modelling, which are scaled with the metallicity, the} lack of a dependence of the O3C2 ratio on the gas-phase metallicity \rev{for subsolar metallicities} is expected. 
\rev{F}or all metallicities, the C/O ratio is constant \rev{in our models, while in galaxies, the C/O ratio would depend on metallicity because the timescales of the enrichment processes are different \citep{MM19, Arata20}}. 
\rev{At supersolar metallicities, the O3C2 ratio increases sharply by two orders of magnitude. We note that only a small fraction of the gas mass in \Ponos~(5.2\%) reaches high metallicities like this. }
\rev{The increase in the ratio as a function of metallicity is caused by the properties of the gas, as it seems to have recently been enriched by core-collapse supernova (cc-SNe) and is generally heated, has low densities, and experiences a strong radiation field.}
\rev{At log$_{10}$(Z/Z$_{\odot}$) $\geq$ 0.2, we find gas with O3C2 $\geq$ 10,  with temperatures of 10$^{4}$ - 2$\cdot$10$^{4}$~K and densities of n$\sim$ 0.1 - 10 cm$^{-3}$. It also experiences strong radiation fields of log(G$_0$) $\sim$ 2.0 - 2.5.}
\rev{We attribute the sharp increase in the O3C2 emission line ratio for supersolar metallicities to the effect of the radiation field and to the low gas densities and high temperatures. From observational results \citep{Cormier15} and theoretical modelling, where individual elements are traced \citep{Arata20}, we would expect lower O3C2 ratios at higher metallicities.}
A direct tracing of carbon and oxygen with their respective yields caused by different enrichment events (e.g. cc-SNe, SN Type Ia, and asymptotic giant branch (AGB) stars) could reveal a significant metallicity dependence. \cite{Arata20} found in their semi-analytical modelling that the O3C2 ratio decreases by two orders of magnitude with increasing metallicity, which could counteract the trend we find when we assumed a fixed metallicity-independent C/O ratio. 
The ratio has a larger scatter as a function of both gas-phase metallicity and radiation field compared to density and temperature ($\leq$~0.5 dex). 
The last column of Fig.~\ref{fig:HistRatio} shows the ratio as a function of the absolute velocity of the gas, divided into inflowing and outflowing components. The outflowing gas shows higher O3C2 ratios than the inflowing gas, which is due to gas heated by SNe feedback, where [OIII] radiation dominates.

\subsection{O3$_{52}$O3$_{88}$}

The [OIII]~$\lambda52\mu$m / [OIII]~$\lambda88\mu$m (O3$_{52}$O3$_{88}$) ratio has been found to be an indicator for the gas density \citep{Dinerstein85,Rubin94, Brauher08, Nagao11, Nagao13, Pereira17, Rigo18, Killi23, Nakazato23}. In contrast to the optical [OIII] lines (e.g. 4363$\AA{}$, 4959$\AA{}$, and 5007$\AA{}$), the FIR [OIII] \rev{line ratio is} indeed less dependent on the electron temperature and more sensitive to the electron density \citep{Dinerstein85}. 

The total O3$_{52}$O3$_{88}$ of \Ponos~is 0.36. \rev{Recent ALMA observations by \cite{Killi23} found a ratio of 0.7 for a galaxy at z = 7, which is higher by a factor of two  than \Ponos. Simulation work by \cite{Nakazato23} found their simulated galaxies at z = 7 to have ratios $\leq 1$, ranging from about 0.8 - 1.0. } 
\rev{Our result, as well as the findings by \cite{Killi23} and \cite{Nakazato23}, is} lower than results for local galaxies \citep{Brauher08} and local planetary nebulae \citep{Dinerstein85}. 
As shown in the second panel of Fig.~\ref{fig:RatioMaps}, there are spatial variations in the map, with generally lower ratios in the inner regions of the disc and merger. The ratios increase radially towards the outskirts. 
In the second row of Fig.~\ref{fig:HistRatio}, the strongest dependence can be seen on the gas density, where the scatter is also very small, which agrees with previous findings \citep{Nagao11,Pereira17}. This is because the two [OIII] lines have slightly different critical densities ($3.6 \cdot 10^{3}$~cm$^{-3}$ for [OIII]52 and $5.1 \cdot 10^{2}$~cm$^{-3}$ for [OIII]88), and thus, their ratio is sensitive to the gas density. Higher densities lead to higher ratios, with an increase of four orders of magnitude from n~=~$10^{-2.5}$~cm$^{-3}$ to n~=~$10^{2.5}$~cm$^{-3}$. The dependence on temperature is weak and degenerate because higher O3$_{52}$O3$_{88}$ ratio values are associated with both low and high temperatures. 
All galaxy components show a completely flat trend of O3$_{52}$O3$_{88}$ as a function of gas metallicity, which is expected because the two lines trace the same element. 
There is a dependence on the radiation field, where we see a rise in the ratio towards higher radiation fields.
The outflowing and inflowing component show a significant difference in ratio as a function of the velocity, where the ratio \rev{of the outflowing gas} decreases by over an order of magnitudes. This indicates that the [OIII]88 line is a stronger tracer of outflows than the [OIII]52 emission line.

\subsection{N2N3}

The [NII]~$\lambda 122 \mu$m / [NIII]~$\lambda 57 \mu$m (N2N3) ratio can be used to study the ionisation parameter of the gas \citep{Nagao11} and to measure the \rev{effective temperature of stars that ionise the gas} \citep{Rubin94,Brauher08}. Singly ionised nitrogen (N$^{+}$) has an ionisation potential of 14.53 eV and a critical density of $3.1 \cdot 10^{2}$~cm$^{-3}$, and thus, [NII] traces areas with a lower temperature in ionised HII regions than doubly ionised nitrogen (N$^{++}$), which has an ionisation potential of 29.60 eV. 

The third panel of Fig.~\ref{fig:RatioMaps} shows that in contrast to the previously discussed ratios, the N2N3 ratio is almost uniform across the source, with an enhancement in small areas of lower-temperature ionised gas (10$^{4}$~K~<~T~<~10$^{4.5}$~K), where [NII] dominates. 
In the third row of Fig.~\ref{fig:HistRatio}, the N2N3 ratio is degenerate with the gas density. The most significant dependence of this ratio is on the gas temperature, where higher temperatures lead to lower ratios. 
The decrease of N2N3 as a function of temperature is steeper for disc and merger components than for the CGM because of the radiation field strength of the CGM is low. 
\rev{As metals in gas promote the cooling, we investigated whether the ability of nitrogen to trace the temperature of the ionised gas could be affected by a change in metallicity. We found no significant differences in the dependence of the N2N3 ratio on the temperature at different metallicities.}

The N2N3 ratio does not depend on the metallicity until supersolar values are reached, where the ratio rapidly drops. This can once again be connected to hot gas that has been \rev{enriched by SNe and experiences a strong radiation field}. This is the same trend as in the corresponding increase of O3C2.
There is also a dependence on the radiation field, with higher ratios for lower radiation fields. This is caused by the lower ionisation energy of [NII] compared to [NIII]. 
We do not detect a significant difference in N2N3 ratio between inflows and outflows.
Due to the weak dependences on density and metallicity, we find that N2N3 can be used as a diagnostic tool \rev{for constraining the ionised gas temperature. We also find a dependence on the radiation field, and thus, on the ionisation of the gas, which agrees with \cite{Nagao11}, who found that N2N3 traces the ionisation parameter.}

\subsection{C2N2}

Like the O3C2 ratio, the [CII]~$\lambda 158 \mu$m~/~[NII]~$\lambda 122 \mu$m (C2N2) ratio uses two emission lines that originate in different phases of the ISM. The [CII] and [NII] lines have ionisation energies of 11.26~eV and 14.53~eV, respectively, which are both close to the ionisation energy of hydrogen (13.6~eV). Previous studies have tested the [CII]~$\lambda 158 \mu$m~/~[NII]~$\lambda 205 \mu$m (or [NII]205/[CII]) ratio. It was proposed to be a tracer of metallicity \citep{Nagao13} and as a tracer for constraining the fraction of [CII] emission that is associated with ionised gas \citep{Decarli14}. It was also suggested scale with the IR luminosity \citep{Cunningham20}.
The ratio of [CII] and [NII]122 has not been well explored so far.

The spatially resolved map of this ratio (bottom left panel of Fig.~\ref{fig:RatioMaps}) shows higher values in dense structures of the disc.
The plots reported in Fig.~\ref{fig:HistRatio} (fourth row) show evidence for a phase transition of C2N2 as a function of density and temperature. In the low-density regime, the ratio has a slight tendency to decrease with increasing density until it reaches a turning point at n = 5~cm$^{-3}$, above which the ratio rapidly increases by two orders of magnitude with density. [CII] has a wide range of critical densities for collisions with electrons, atoms, or molecules (5~cm$^{-3}$ up to 10$^{3}$~cm$^{-3}$), and the C2N2 ratio starts to increase when the minimum critical density is reached. The same behaviour can be seen in the temperature, where the ratio quickly drops off by two orders of magnitude until it reaches a temperature slightly above 10$^{4}$~K, after which it remains roughly constant with temperature. Therefore, denser and colder regions of the ISM are more sensitive to the C2N2 ratio than diffuse gas, which is also visible in Fig.~\ref{fig:RatioMaps}. The turning point in both temperature and density shows evidence for a phase transition from ionised HII regions to neutral PDR regions. 
Therefore, we find that the C2N2 ratio could be used to infer the gas density and temperature of PDR regions, as it seems to be most sensitive in this regime. 
On the other hand, the C2N2 ratio is insensitive to the gas-phase metallicity. This insensitivity would not significantly change if we were able to directly trace the abundances of elements, as the C/N ratio has been found to be approximately constant over a wide range of metallicities \citep{Garnett99, Berg16, Pereira17}.
The ratio is also only slightly sensitive to the radiation field, where a large scatter within the disc component can be seen. In the merger and the disc, the ratio increases towards lower radiation fields, which corresponds to shielded, denser gas. There is a general weak trend for lower ratios in higher radiation fields. The C2N2 ratio remains constant between the outflowing and inflowing gas. 

\subsection{O3N3}

The [OIII]~$\lambda 88 \mu$m / [NIII]~$\lambda 57 \mu$m (O3N3) ratio is a candidate for tracing the gas metallicity because it is a good proxy for the N/O abundance ratio \citep{Nagao11, Pereira17}. The two lines have similar ionisation energies and thus trace the same gas in HII regions, which is important for metallicity calibrations \citep{MD23}. 
To account for the dependence of the [OIII] lines on the electron density of the gas, the O3N3 ratio often combines the [OIII]88 and the [OIII]52 line \citep{Pereira17}. In our analysis, we found that including the [OIII]52 line does not change our results, and thus, we focused on the ratio that only included the [OIII]88 line (which is easier to observe from the ground). 
As discussed above for the O3C2 ratio, the assumption of constant abundance ratios in our modelling limits our capability of exploring the true metallicity dependences. 

The second panel in the second row of Fig.~\ref{fig:RatioMaps} shows the O3N3 ratio map.
The two lines originate from similar phases of the gas, and the spatial variation of the ratio is therefore small. The O3N3 ratio is higher in the inner regions of the disc. Small areas with highly increased ratios within the ISM correspond to hot and diffuse regions, such as HII regions and SNe sites, where [OIII] dominates.   
The bottom row of Fig.~\ref{fig:HistRatio} shows a strong dependence of the ratio on the gas density, whereby the ratio decreases by over two orders of magnitude within the explored density range. The trend with temperature shows a similar behaviour, where gas with lower temperatures also has a lower ratio, and it flattens towards higher temperatures.  
The radiation field seems to be the strongest dependence we find, where high-radiation fields lead to a low ratio that rises by two orders of magnitude.
Interestingly, the ratio of the outflowing and inflowing gas is different, whereby the O3N3 ratio increases by an order of magnitude as a function of velocity for the former.

\section{Conclusions}

We investigated FIR emission line ratios in the context of a cosmological zoom-in simulation of a high-redshift galaxy to asses the information provided by these ratios about the ISM properties of the galaxy. Our main results are listed below.

\begin{itemize}
    \item We discovered that C2N2 can be used as a tracer to infer the density and temperature of PDR regions.
    \item O3O3 is confirmed to be a good tracer for the gas-phase density, although we also find a dependence on the incident radiation field. 
    \item We find a significant difference of O3C2, O3O3, and O3N3 in in- and outflowing gas, suggesting that these ratios could be used to analyse high-velocity components. [OIII]88 $\mu$m is sensitive to high-velocity outflowing gas.
    \item N2N3 is \rev{found} to be a good tracer of the temperature \rev{of ionised gas and is confirmed to trace the radiation field} in the ISM of galaxies.
\end{itemize}


\begin{acknowledgements}
This project has received funding from the European Union’s Horizon 2020 research and innovation program under grant agreement No 951815 (AtLAST). The simulations were performed using the resources from the National Infrastructure for High Performance Computing and Data Storage in Norway, UNINETT Sigma2, allocated to Project NN9477K. \rev{We thank the anonymous referee for providing insightful comments that helped us improve the analysis and interpretation of the results.}
We acknowledge usage of the Python programming language \citep{van1995python,python3}, Matplotlib \citep{Hunter07}, NumPy \citep{Walt2011}, and Pynbody \citep{Pynbody}.
\end{acknowledgements}

%
%
\bibliography{ponos}

\begin{thebibliography}{58}
\expandafter\ifx\csname natexlab\endcsname\relax\def\natexlab#1{#1}\fi

\bibitem[{{Arata} {et~al.}(2020){Arata}, {Yajima}, {Nagamine}, {Abe}, \& {Khochfar}}]{Arata20}
{Arata}, S., {Yajima}, H., {Nagamine}, K., {Abe}, M., \& {Khochfar}, S. 2020, \mnras, 498, 5541

\bibitem[{{Berg} {et~al.}(2016){Berg}, {Skillman}, {Henry}, {Erb}, \& {Carigi}}]{Berg16}
{Berg}, D.~A., {Skillman}, E.~D., {Henry}, R. B.~C., {Erb}, D.~K., \& {Carigi}, L. 2016, \apj, 827, 126

\bibitem[{{Brauher} {et~al.}(2008){Brauher}, {Dale}, \& {Helou}}]{Brauher08}
{Brauher}, J.~R., {Dale}, D.~A., \& {Helou}, G. 2008, \apjs, 178, 280

\bibitem[{{Carniani} {et~al.}(2020){Carniani}, {Ferrara}, {Maiolino}, {Castellano}, {Gallerani}, {Fontana}, {Kohandel}, {Lupi}, {Pallottini}, {Pentericci}, {Vallini}, \& {Vanzella}}]{Carniani20}
{Carniani}, S., {Ferrara}, A., {Maiolino}, R., {et~al.} 2020, \mnras, 499, 5136

\bibitem[{{Carniani} {et~al.}(2018){Carniani}, {Maiolino}, {Amorin}, {Pentericci}, {Pallottini}, {Ferrara}, {Willott}, {Smit}, {Matthee}, {Sobral}, {Santini}, {Castellano}, {De Barros}, {Fontana}, {Grazian}, \& {Guaita}}]{Carniani18}
{Carniani}, S., {Maiolino}, R., {Amorin}, R., {et~al.} 2018, \mnras, 478, 1170

\bibitem[{{Cormier} {et~al.}(2015){Cormier}, {Madden}, {Lebouteiller}, {Abel}, {Hony}, {Galliano}, {R{\'e}my-Ruyer}, {Bigiel}, {Baes}, {Boselli}, {Chevance}, {Cooray}, {De Looze}, {Doublier}, {Galametz}, {Hughes}, {Karczewski}, {Lee}, {Lu}, \& {Spinoglio}}]{Cormier15}
{Cormier}, D., {Madden}, S.~C., {Lebouteiller}, V., {et~al.} 2015, \aap, 578, A53

\bibitem[{{Cunningham} {et~al.}(2020){Cunningham}, {Chapman}, {Aravena}, {De Breuck}, {B{\'e}thermin}, {Chen}, {Dong}, {Gonzalez}, {Greve}, {Litke}, {Ma}, {Malkan}, {Marrone}, {Miller}, {Phadke}, {Reuter}, {Rotermund}, {Spilker}, {Stark}, {Strandet}, {Vieira}, \& {Wei{\ss}}}]{Cunningham20}
{Cunningham}, D.~J.~M., {Chapman}, S.~C., {Aravena}, M., {et~al.} 2020, \mnras, 494, 4090

\bibitem[{Decarli {et~al.}(2014)Decarli, Walter, Carilli, Bertoldi, Cox, Ferkinhoff, Groves, Maiolino, Neri, Riechers, \& Weiss}]{Decarli14}
Decarli, R., Walter, F., Carilli, C., {et~al.} 2014, The Astrophysical Journal Letters, 782, L17

\bibitem[{{Decataldo} {et~al.}(2020){Decataldo}, {Lupi}, {Ferrara}, {Pallottini}, \& {Fumagalli}}]{Decataldo20}
{Decataldo}, D., {Lupi}, A., {Ferrara}, A., {Pallottini}, A., \& {Fumagalli}, M. 2020, \mnras, 497, 4718

\bibitem[{{Decataldo} {et~al.}(2019){Decataldo}, {Pallottini}, {Ferrara}, {Vallini}, \& {Gallerani}}]{Decataldo19}
{Decataldo}, D., {Pallottini}, A., {Ferrara}, A., {Vallini}, L., \& {Gallerani}, S. 2019, \mnras, 487, 3377

\bibitem[{{Dinerstein} {et~al.}(1985){Dinerstein}, {Lester}, \& {Werner}}]{Dinerstein85}
{Dinerstein}, H.~L., {Lester}, D.~F., \& {Werner}, M.~W. 1985, \apj, 291, 561

\bibitem[{{Ferland} {et~al.}(2017){Ferland}, {Chatzikos}, {Guzm{\'a}n}, {Lykins}, {van Hoof}, {Williams}, {Abel}, {Badnell}, {Keenan}, {Porter}, \& {Stancil}}]{Cloudy}
{Ferland}, G.~J., {Chatzikos}, M., {Guzm{\'a}n}, F., {et~al.} 2017, \rmxaa, 53, 385

\bibitem[{{Fiacconi} {et~al.}(2017){Fiacconi}, {Mayer}, {Madau}, {Lupi}, {Dotti}, \& {Haardt}}]{Fiacconi17}
{Fiacconi}, D., {Mayer}, L., {Madau}, P., {et~al.} 2017, \mnras, 467, 4080

\bibitem[{{Fujimoto} {et~al.}(2022){Fujimoto}, {Ouchi}, {Nakajima}, {Harikane}, {Isobe}, {Brammer}, {Oguri}, {Gim{\'e}nez-Arteaga}, {Heintz}, {Kokorev}, {Bauer}, {Ferrara}, {Kojima}, {Lagos}, {Laura}, {Schaerer}, {Shimasaku}, {Hatsukade}, {Kohno}, {Sun}, {Valentino}, {Watson}, {Fudamoto}, {Inoue}, {Gonz{\'a}lez-L{\'o}pez}, {Koekemoer}, {Knudsen}, {Lee}, {Magdis}, {Richard}, {Strait}, {Sugahara}, {Tamura}, {Toft}, {Umehata}, \& {Walth}}]{Fujimoto22}
{Fujimoto}, S., {Ouchi}, M., {Nakajima}, K., {et~al.} 2022, arXiv e-prints, arXiv:2212.06863

\bibitem[{{Garnett} {et~al.}(1999){Garnett}, {Shields}, {Peimbert}, {Torres-Peimbert}, {Skillman}, {Dufour}, {Terlevich}, \& {Terlevich}}]{Garnett99}
{Garnett}, D.~R., {Shields}, G.~A., {Peimbert}, M., {et~al.} 1999, \apj, 513, 168

\bibitem[{{Grassi} {et~al.}(2014){Grassi}, {Bovino}, {Schleicher}, {Prieto}, {Seifried}, {Simoncini}, \& {Gianturco}}]{KROME}
{Grassi}, T., {Bovino}, S., {Schleicher}, D.~R.~G., {et~al.} 2014, \mnras, 439, 2386

\bibitem[{{Haardt} \& {Madau}(2012)}]{HM12}
{Haardt}, F. \& {Madau}, P. 2012, \apj, 746, 125

\bibitem[{{Habing}(1968)}]{Habing68}
{Habing}, H.~J. 1968, \bain, 19, 421

\bibitem[{{Harikane} {et~al.}(2020){Harikane}, {Ouchi}, {Inoue}, {Matsuoka}, {Tamura}, {Bakx}, {Fujimoto}, {Moriwaki}, {Ono}, {Nagao}, {Tadaki}, {Kojima}, {Shibuya}, {Egami}, {Ferrara}, {Gallerani}, {Hashimoto}, {Kohno}, {Matsuda}, {Matsuo}, {Pallottini}, {Sugahara}, \& {Vallini}}]{Harikane20}
{Harikane}, Y., {Ouchi}, M., {Inoue}, A.~K., {et~al.} 2020, \apj, 896, 93

\bibitem[{Hashimoto {et~al.}(2019)Hashimoto, Inoue, Mawatari, Tamura, Matsuo, Furusawa, Harikane, Shibuya, Knudsen, Kohno, Ono, Zackrisson, Okamoto, Kashikawa, Oesch, Ouchi, Ota, Shimizu, Taniguchi, Umehata, \& Watson}]{Hashimoto19}
Hashimoto, T., Inoue, A.~K., Mawatari, K., {et~al.} 2019, Publications of the Astronomical Society of Japan, 71, 71

\bibitem[{{Hinshaw} {et~al.}(2013){Hinshaw}, {Larson}, {Komatsu}, {Spergel}, {Bennett}, {Dunkley}, {Nolta}, {Halpern}, {Hill}, {Odegard}, {Page}, {Smith}, {Weiland}, {Gold}, {Jarosik}, {Kogut}, {Limon}, {Meyer}, {Tucker}, {Wollack}, \& {Wright}}]{Hinshaw13}
{Hinshaw}, G., {Larson}, D., {Komatsu}, E., {et~al.} 2013, \apjs, 208, 19

\bibitem[{Hunter(2007)}]{Hunter07}
Hunter, J.~D. 2007, Computing in Science \& Engineering, 9, 90

\bibitem[{{Inoue} {et~al.}(2016){Inoue}, {Tamura}, {Matsuo}, {Mawatari}, {Shimizu}, {Shibuya}, {Ota}, {Yoshida}, {Zackrisson}, {Kashikawa}, {Kohno}, {Umehata}, {Hatsukade}, {Iye}, {Matsuda}, {Okamoto}, \& {Yamaguchi}}]{Inoue16}
{Inoue}, A.~K., {Tamura}, Y., {Matsuo}, H., {et~al.} 2016, Science, 352, 1559

\bibitem[{{Katz} {et~al.}(2019){Katz}, {Galligan}, {Kimm}, {Rosdahl}, {Haehnelt}, {Blaizot}, {Devriendt}, {Slyz}, {Laporte}, \& {Ellis}}]{Katz19}
{Katz}, H., {Galligan}, T.~P., {Kimm}, T., {et~al.} 2019, \mnras, 487, 5902

\bibitem[{{Katz} {et~al.}(2022){Katz}, {Rosdahl}, {Kimm}, {Garel}, {Blaizot}, {Haehnelt}, {Michel-Dansac}, {Martin-Alvarez}, {Devriendt}, {Slyz}, {Teyssier}, {Ocvirk}, {Laporte}, \& {Ellis}}]{Katz22}
{Katz}, H., {Rosdahl}, J., {Kimm}, T., {et~al.} 2022, \mnras, 510, 5603

\bibitem[{{Killi} {et~al.}(2023){Killi}, {Watson}, {Fujimoto}, {Akins}, {Knudsen}, {Richard}, {Harikane}, {Rigopoulou}, {Rizzo}, {Ginolfi}, {Popping}, \& {Kokorev}}]{Killi23}
{Killi}, M., {Watson}, D., {Fujimoto}, S., {et~al.} 2023, \mnras, 521, 2526

\bibitem[{{Komatsu} {et~al.}(2011){Komatsu}, {Smith}, {Dunkley}, {Bennett}, {Gold}, {Hinshaw}, {Jarosik}, {Larson}, {Nolta}, {Page}, {Spergel}, {Halpern}, {Hill}, {Kogut}, {Limon}, {Meyer}, {Odegard}, {Tucker}, {Weiland}, {Wollack}, \& {Wright}}]{Komatsu11}
{Komatsu}, E., {Smith}, K.~M., {Dunkley}, J., {et~al.} 2011, \apjs, 192, 18

\bibitem[{{Kumari} {et~al.}(2023){Kumari}, {Smit}, {Leitherer}, {Witstok}, {Irwin}, {Sirianni}, \& {Aloisi}}]{Kumari23}
{Kumari}, N., {Smit}, R., {Leitherer}, C., {et~al.} 2023, arXiv e-prints, arXiv:2307.00059

\bibitem[{{Kumari} {et~al.}(2024){Kumari}, {Smit}, {Leitherer}, {Witstok}, {Irwin}, {Sirianni}, \& {Aloisi}}]{Kumari24}
{Kumari}, N., {Smit}, R., {Leitherer}, C., {et~al.} 2024, \mnras, 529, 781

\bibitem[{{Leitherer} {et~al.}(1999){Leitherer}, {Schaerer}, {Goldader}, {Delgado}, {Robert}, {Kune}, {de Mello}, {Devost}, \& {Heckman}}]{SB99}
{Leitherer}, C., {Schaerer}, D., {Goldader}, J.~D., {et~al.} 1999, \apjs, 123, 3

\bibitem[{{Lupi} {et~al.}(2020){Lupi}, {Pallottini}, {Ferrara}, {Bovino}, {Carniani}, \& {Vallini}}]{Lupi2020}
{Lupi}, A., {Pallottini}, A., {Ferrara}, A., {et~al.} 2020, \mnras, 496, 5160

\bibitem[{{Maiolino} \& {Mannucci}(2019)}]{MM19}
{Maiolino}, R. \& {Mannucci}, F. 2019, \aapr, 27, 3

\bibitem[{{McKee} \& {Ostriker}(1977)}]{McKee77}
{McKee}, C.~F. \& {Ostriker}, J.~P. 1977, \apj, 218, 148

\bibitem[{{M{\'e}ndez-Delgado} {et~al.}(2023){M{\'e}ndez-Delgado}, {Esteban}, {Garc{\'\i}a-Rojas}, {Kreckel}, \& {Peimbert}}]{MD23}
{M{\'e}ndez-Delgado}, J.~E., {Esteban}, C., {Garc{\'\i}a-Rojas}, J., {Kreckel}, K., \& {Peimbert}, M. 2023, \nat, 618, 249

\bibitem[{Mroczkowski {et~al.}(2024)Mroczkowski, Gallardo, Timpe, Kiselev, Groh, Kaercher, Reichert, Cicone, Puddu, dit Bonclaude, Bok, Dahl, Macintosh, Dicker, Viole, Sartori, Venegas, Zeyringer, Niemack, Poppi, Olguin, Hatziminaoglou, Breuck, Klaassen, Montenegro-Montes, \& Zimmerer}]{Mroczkowski2024}
Mroczkowski, T., Gallardo, P.~A., Timpe, M., {et~al.} 2024, arXiv e-prints, arXiv:2402.18645

\bibitem[{{Nagao} {et~al.}(2013){Nagao}, {Maiolino}, {De Breuck}, {Caselli}, {Hatsukade}, \& {Saigo}}]{Nagao13}
{Nagao}, T., {Maiolino}, R., {De Breuck}, C., {et~al.} 2013, in Astronomical Society of the Pacific Conference Series, Vol. 476, New Trends in Radio Astronomy in the ALMA Era: The 30th Anniversary of Nobeyama Radio Observatory, ed. R.~{Kawabe}, N.~{Kuno}, \& S.~{Yamamoto}, 29

\bibitem[{{Nagao} {et~al.}(2011){Nagao}, {Maiolino}, {Marconi}, \& {Matsuhara}}]{Nagao11}
{Nagao}, T., {Maiolino}, R., {Marconi}, A., \& {Matsuhara}, H. 2011, \aap, 526, A149

\bibitem[{{Nakazato} {et~al.}(2023){Nakazato}, {Yoshida}, \& {Ceverino}}]{Nakazato23}
{Nakazato}, Y., {Yoshida}, N., \& {Ceverino}, D. 2023, \apj, 953, 140

\bibitem[{Osterbrock \& Ferland(2006)}]{Osterbrock06}
Osterbrock, D.~E. \& Ferland, G.~J. 2006, Astrophysics of gaseous nebulae and active galactic nuclei (University Science Books)

\bibitem[{{Pallottini} {et~al.}(2017){Pallottini}, {Ferrara}, {Bovino}, {Vallini}, {Gallerani}, {Maiolino}, \& {Salvadori}}]{Pallottini17}
{Pallottini}, A., {Ferrara}, A., {Bovino}, S., {et~al.} 2017, \mnras, 471, 4128

\bibitem[{{Pallottini} {et~al.}(2019){Pallottini}, {Ferrara}, {Decataldo}, {Gallerani}, {Vallini}, {Carniani}, {Behrens}, {Kohandel}, \& {Salvadori}}]{Pallottini19}
{Pallottini}, A., {Ferrara}, A., {Decataldo}, D., {et~al.} 2019, \mnras, 487, 1689

\bibitem[{{Pallottini} {et~al.}(2022){Pallottini}, {Ferrara}, {Gallerani}, {Behrens}, {Kohandel}, {Carniani}, {Vallini}, {Salvadori}, {Gelli}, {Sommovigo}, {D'Odorico}, {Di Mascia}, \& {Pizzati}}]{Pallottini22}
{Pallottini}, A., {Ferrara}, A., {Gallerani}, S., {et~al.} 2022, \mnras, 513, 5621

\bibitem[{Pereira-Santaella {et~al.}(2017)Pereira-Santaella, Rigopoulou, Farrah, Lebouteiller, \& Li}]{Pereira17}
Pereira-Santaella, M., Rigopoulou, D., Farrah, D., Lebouteiller, V., \& Li, J. 2017, Monthly Notices of the Royal Astronomical Society, 470, 1218

\bibitem[{{Pontzen} {et~al.}(2013){Pontzen}, {Ro{\v{s}}kar}, {Stinson}, \& {Woods}}]{Pynbody}
{Pontzen}, A., {Ro{\v{s}}kar}, R., {Stinson}, G., \& {Woods}, R. 2013, {pynbody: N-Body/SPH analysis for python}

\bibitem[{{Ramos Padilla} {et~al.}(2023){Ramos Padilla}, {Wang}, {van der Tak}, \& {Trager}}]{RP23}
{Ramos Padilla}, A.~F., {Wang}, L., {van der Tak}, F.~F.~S., \& {Trager}, S.~C. 2023, \aap, 679, A131

\bibitem[{{Rigopoulou} {et~al.}(2018){Rigopoulou}, {Pereira-Santaella}, {Magdis}, {Cooray}, {Farrah}, {Marques-Chaves}, {Perez-Fournon}, \& {Riechers}}]{Rigo18}
{Rigopoulou}, D., {Pereira-Santaella}, M., {Magdis}, G.~E., {et~al.} 2018, \mnras, 473, 20

\bibitem[{{Rosdahl} {et~al.}(2013){Rosdahl}, {Blaizot}, {Aubert}, {Stranex}, \& {Teyssier}}]{RamsesRT}
{Rosdahl}, J., {Blaizot}, J., {Aubert}, D., {Stranex}, T., \& {Teyssier}, R. 2013, \mnras, 436, 2188

\bibitem[{{Rubin} {et~al.}(1994){Rubin}, {Simpson}, {Lord}, {Colgan}, {Erickson}, \& {Haas}}]{Rubin94}
{Rubin}, R.~H., {Simpson}, J.~P., {Lord}, S.~D., {et~al.} 1994, \apj, 420, 772

\bibitem[{{Schimek} {et~al.}(2024){Schimek}, {Decataldo}, {Shen}, {Cicone}, {Baumschlager}, {van Kampen}, {Klaassen}, {Madau}, {Di Mascolo}, {Mayer}, {Montoya Arroyave}, {Mroczkowski}, \& {Warraich}}]{Schimek23}
{Schimek}, A., {Decataldo}, D., {Shen}, S., {et~al.} 2024, \aap, 682, A98

\bibitem[{Stacey {et~al.}(1991)Stacey, Townes, Poglitsch, Madden, Jackson, Herrmann, Genzel, \& Geis}]{Stacey91}
Stacey, G., Townes, C., Poglitsch, A., {et~al.} 1991, The Astrophysical Journal, 382, L37

\bibitem[{{Teyssier}(2002)}]{Ramses}
{Teyssier}, R. 2002, \aap, 385, 337

\bibitem[{{Vallini} {et~al.}(2021){Vallini}, {Ferrara}, {Pallottini}, {Carniani}, \& {Gallerani}}]{Vallini21}
{Vallini}, L., {Ferrara}, A., {Pallottini}, A., {Carniani}, S., \& {Gallerani}, S. 2021, \mnras, 505, 5543

\bibitem[{{Vallini} {et~al.}(2015){Vallini}, {Gallerani}, {Ferrara}, {Pallottini}, \& {Yue}}]{Vallini15}
{Vallini}, L., {Gallerani}, S., {Ferrara}, A., {Pallottini}, A., \& {Yue}, B. 2015, \apj, 813, 36

\bibitem[{{van der Walt} {et~al.}(2011){van der Walt}, {Colbert}, \& {Varoquaux}}]{Walt2011}
{van der Walt}, S., {Colbert}, S.~C., \& {Varoquaux}, G. 2011, Computing in Science and Engineering, 13, 22

\bibitem[{Van~Rossum \& Drake(2009)}]{python3}
Van~Rossum, G. \& Drake, F.~L. 2009, Python 3 Reference Manual (Scotts Valley, CA: CreateSpace)

\bibitem[{Van~Rossum \& Drake~Jr(1995)}]{van1995python}
Van~Rossum, G. \& Drake~Jr, F.~L. 1995, Python reference manual (Centrum voor Wiskunde en Informatica Amsterdam)

\bibitem[{{Wadsley} {et~al.}(2004){Wadsley}, {Stadel}, \& {Quinn}}]{Wadsley04}
{Wadsley}, J.~W., {Stadel}, J., \& {Quinn}, T. 2004, \na, 9, 137

\bibitem[{{Witstok} {et~al.}(2022){Witstok}, {Smit}, {Maiolino}, {Kumari}, {Aravena}, {Boogaard}, {Bouwens}, {Carniani}, {Hodge}, {Jones}, {Stefanon}, {van der Werf}, \& {Schouws}}]{Witstok22}
{Witstok}, J., {Smit}, R., {Maiolino}, R., {et~al.} 2022, \mnras, 515, 1751

\end{thebibliography}
\bibliographystyle{aa}

\begin{appendix} \label{app}

\section{High metallicity gas in O3C2} \label{app:O3C2}

\begin{figure}
\centering
   \includegraphics[width = \columnwidth]{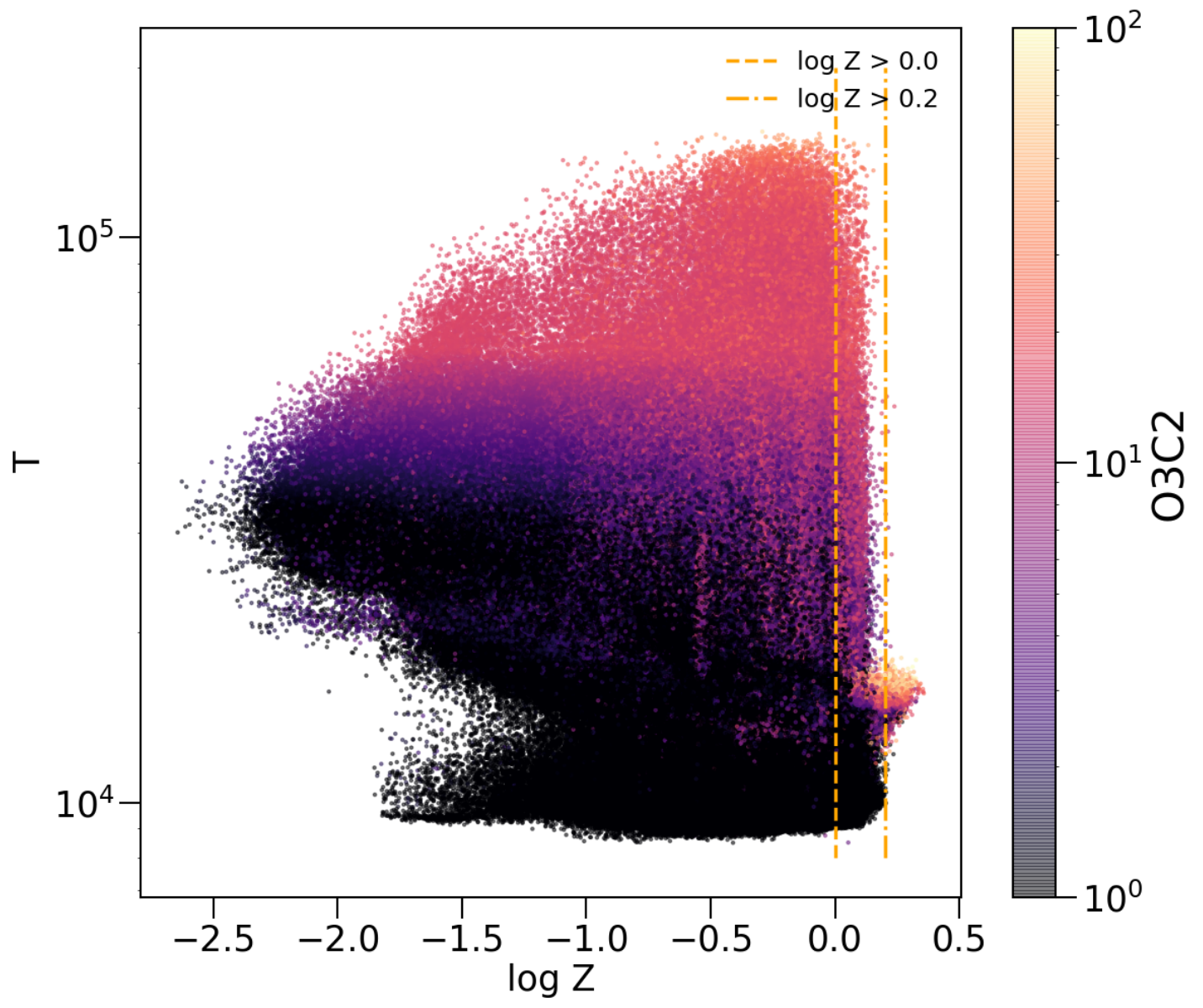} 
      \caption{Phase diagram of gas phase metallicity and gas temperature for gas emitting both in [CII] and [OIII], colour coded by the O3C2 ratio. The vertical dashed line marks solar metallicity, while the dash-dotted line marks log$_{10}$(Z/Z$_{\odot}$ = 0.2).}
         \label{fig:tlogz}
\end{figure}

We investigate the reason behind the increase of the O3C2 ratio at super-solar metallicities in more detail. \autoref{fig:tlogz} shows the phase diagram of the gas emitting in both [CII] and [OIII], with the metallicity on the x-axis and the temperature on the y-axis. As can be seen there is a wide range of temperatures at high metallicity, reaching above 10$^{5}$~K. This high temperatures gas also shows high emission line ratios. Additionally, there is a small region within the phase diagram with very high metallicities, that resides within a small temperature range of around 2$\cdot$10$^{4}$~K, which also displays very high ratios. This gas experiences a high FUV radiation field (log$_{10}$(G$_{0}$) $\sim$ 2 - 2.5) and drives the high metallicity ratio in Fig.~\ref{fig:HistRatio}.

\section{Metallicity effects on the N2N3 ratio} \label{app:n2n3}

\begin{figure}
\centering
   \includegraphics[width = \columnwidth]{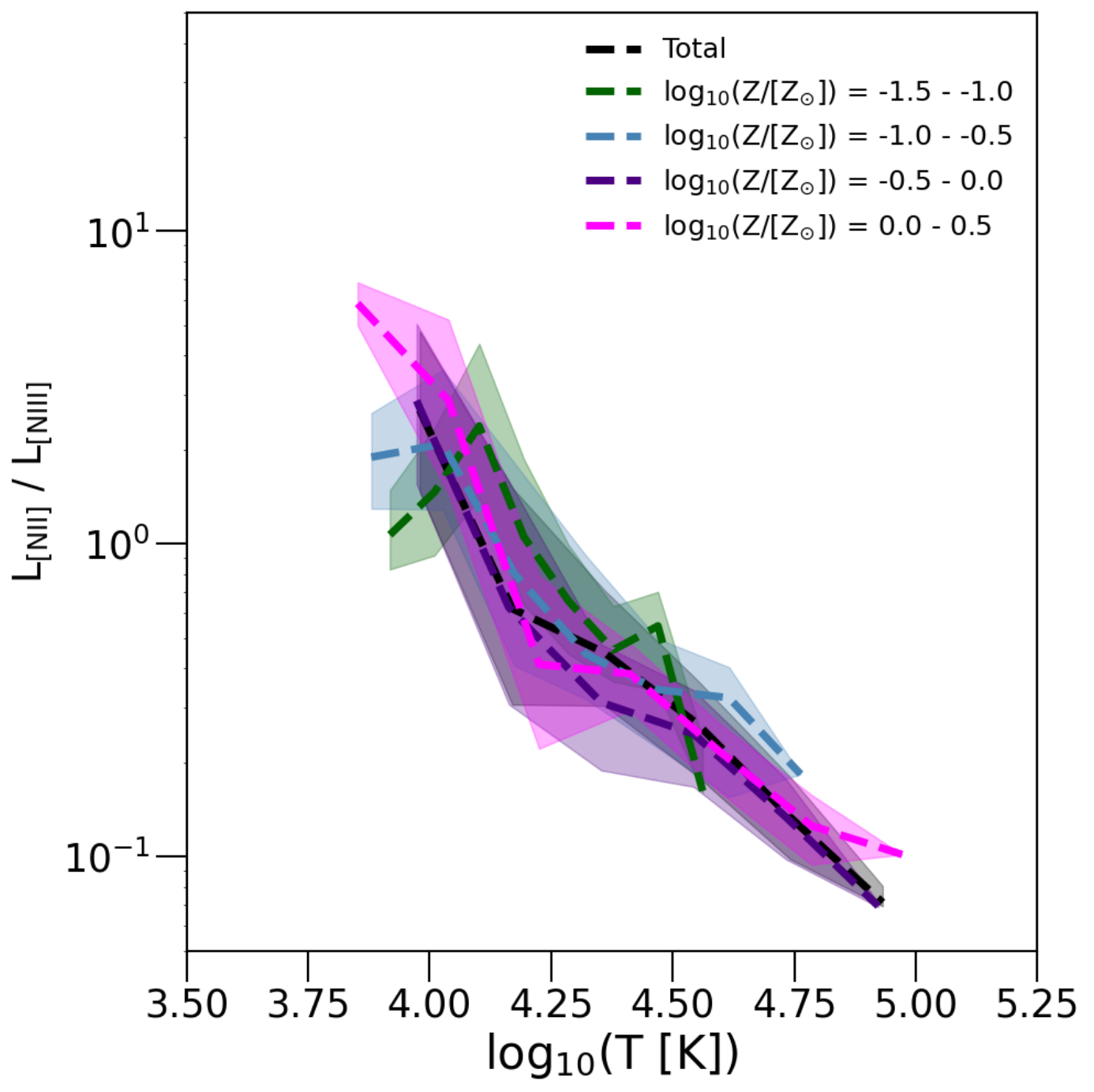} 
      \caption{N2N3 luminosity ratio as a function of temperature, showing the total trend and the gas divided into different metallicity bins.}
         \label{fig:n2n3}
\end{figure}

As metallicity affects the cooling of the gas, it could influence the ability of nitrogen to trace the temperature of the ionised gas.  To investigate this in more detail, we divided the gas of the main disc and the merger component into metallicity bins, to study variations of this ratio  as a function of temperature for different metallicities. This can be seen in Fig.~\ref{fig:n2n3}, where the N2N3 luminosity ratio is plotted as a function of temperature for the different metallicity bins. The plot shows that we see the same steep decreasing trend with temperature for all metallicity bins. We do not see a significant difference in the dependency based on the metallicity of the gas.

\end{appendix}

\end{document}